\documentclass[aps,prl,twocolumn,superscriptaddress,groupedaddress]{revtex4}  % for review and submission
\pdfoutput=1
\usepackage{graphicx}  % needed for figures
\usepackage{dcolumn}   % needed for some tables
\usepackage{bm}        % for math
\usepackage{amssymb}   % for math

\begin{document}

% The following information is for internal review, please remove them for submission
%\widetext
%\leftline{Version 2 as of \today}
%\leftline{Primary authors: Thomas Scheidl, Eric Wille and Rupert Ursin}
%\leftline{To be submitted to Physical Review Letters}
%\leftline{Comment to {\tt rupert.ursin@univie.ac.at}}
%\centerline{\em INTERNAL DOCUMENT -- NOT FOR PUBLIC DISTRIBUTION}

% the following line is for submission, including submission to the arXiv!!
%\hspace{5.2in} \mbox{Fermilab-Pub-04/xxx-E}

%\title{A proposal for performing quantum optics experiments using infrastructure \\onboard the International Space Station}
%\title{A proposal for quantum optics experiment to the ISS}
\title{Quantum optics experiments to the International Space Station ISS: a proposal}

%\input author_list.tex       % D0 authors (remove the first 3 lines
                             % of this file prior to submission, they
                             % contain a time stamp for the authorlist)
                             % (includes institutions and visitors)
\date{\today}
\author{Thomas Scheidl}
\affiliation{Institute for Quantum Optics and Quantum Information, Austrian Academy of Sciences, Boltzmanng. 3, 1090 Vienna, Austria}
\author{Eric Wille}
\affiliation{European Space Agency - ESA/ESTEC Keplerlaan 1, 2201 AZ Noordwijk ZH, The Netherlands}
\author{Rupert Ursin}
\email{Rupert.Ursin@oeaw.ac.at}
\affiliation{Institute for Quantum Optics and Quantum Information, Austrian Academy of Sciences, Boltzmanng. 3, 1090 Vienna, Austria}

\begin{abstract}
We propose performing quantum optics experiments in an ground-to-space scenario using the International Space Station, which is equipped with a glass viewing window and a photographer's lens mounted on a motorized camera pod. A dedicated small add-on module with single-photon detection, time-tagging and classical communication capabilities would enable us to perform the first-ever quantum optics experiments in space. We present preliminary design concepts for the ground and flight segments and study the feasibility of the intended mission scenario.
\end{abstract}

%\pacs{}
\maketitle

%\section{\label{sec:level1}First-level heading}
% sections are not used for PRL papers

Quantum theory was originally developed to describe the smallest entities in physics. Later it turned out, that it also makes fascinating predictions over macroscopic distances. Establishing quantum technology in space enables quantum systems to become available as a resource for quantum physics experiments on a global scale and beyond. The successful implementation of such experiments, which are based on the transmission and detection of single photons or entangled photon pairs, would validate the key technology of a quantum transceiver, involving an entangled photon source, a faint laser pulse source and a single photon counting module. This would represent the first-ever demonstration of a fundamental quantum test (e.g. a Bell-type experiment \cite{zeilinger99b}) and of a quantum communication application (e.g. quantum key distribution, QKD \cite{scarani08a}) in space. In this work, we propose a series of experiments with photons, making use of pre-existing infrastructure onboard the International Space Station (ISS), and also of a dedicated small quantum optics payload based solely on state-of-the-art optical and electronic components. The proposed experiments involve the distribution of entangled photon pairs and faint laser pulses from ground to space. In the following we summarize the relevant quantum communication experiments, present a mission and a preliminary design concept for the ground and the flight segments, and study its feasibility.

The distributing of polarization entangled photon pairs \cite{ursin07} and faint laser pulse decoy-states \cite{Schmitt07} through the atmosphere has already been demonstrated over a terrestrial horizontal link of up to $144$ km between the Canary Islands of La Palma and Tenerife. Recently, successful free-space teleportation has also been demonstrated over $143$ km \cite{Ma12a}. The distribution of time-bin entanglement inside a laboratory \cite{stucki2009a} was shown over a coiled $250$ km long fiber. Moreover, a single photon down-link was emulated using a retro-reflector attached to a satellite reported in Ref. \cite{Villoresi08}.

In order to experimentally test the limits of quantum theory, study the interplay between gravitation and entanglement \cite{ralph06} and to eventually establish a worldwide quantum communication network, it is important to significantly expand the distances for distributing quantum systems beyond the capabilities of terrestrial experiments. Due to transmission losses in fiber based quantum channels and detector dark counts, earth based quantum communication experiments are limited to a distance on the order of $100$ km \cite{Waks02}. One approach for bridging distances on a global scale is the implementation of quantum repeaters, which, however, are in the early stages of development \cite{Rauschenbeutel00a}. Another approach is using free-space links involving satellites.
\begin{figure}[ht]
                \centering
  \includegraphics[width=0.49\textwidth]{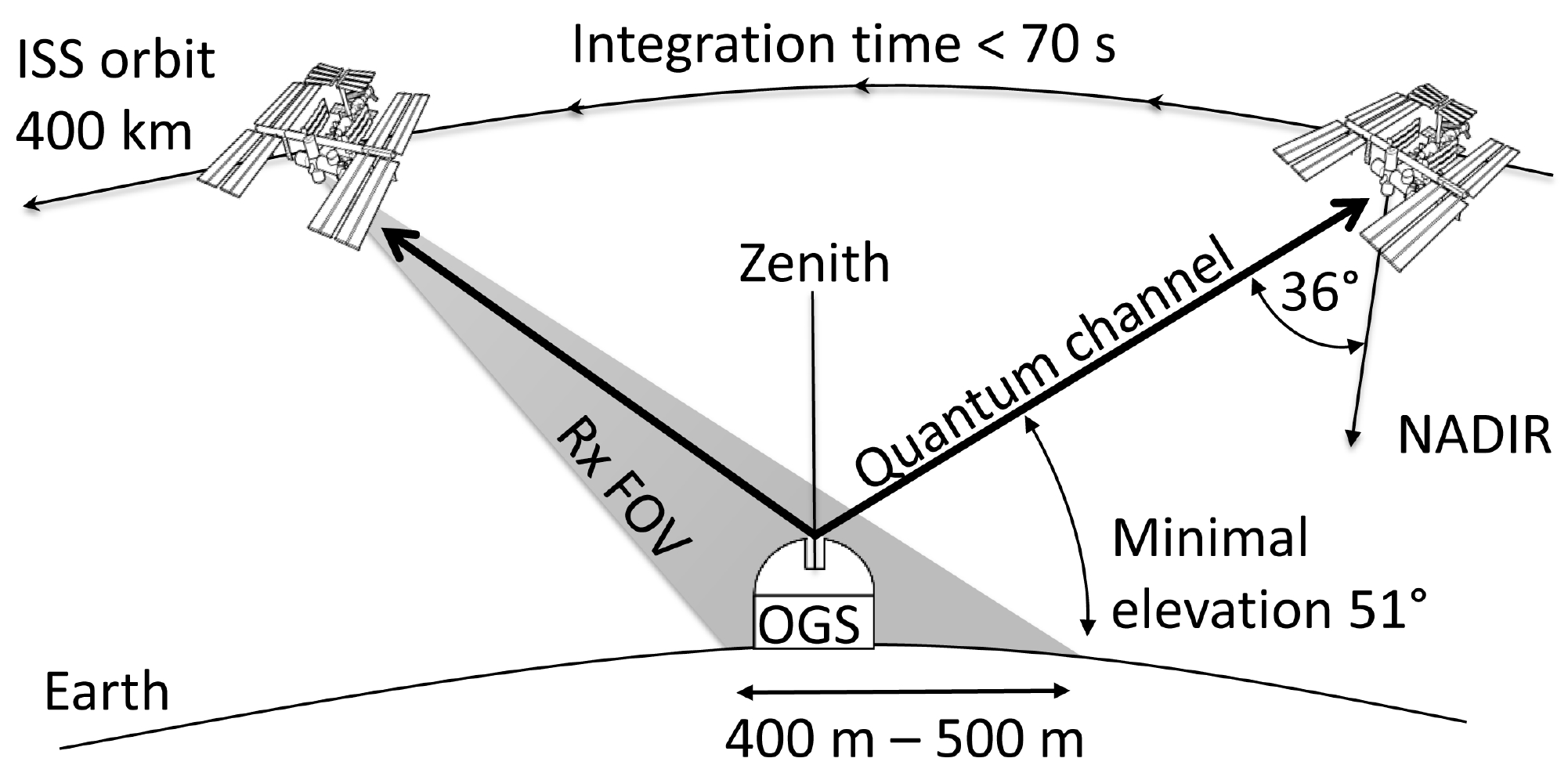}
                  \caption{The orbital pass of the ISS over an optical ground station could be used for quantum communication from inside the Cupola Module, as long as the OGS is not more than $36^\circ$ off the NADIR direction.} %The field-of-vie (FOV) of the single photon detectors determines the area on ground from which light is collected. Calculations show that this is in the order of a few hundred meters. Within such an area, most of the optical ground stations are completely separated from any artificial light source (Lustbühel, Tenerife, ...). Bottom: The angle between the NightPOD's optical axis and the edges of the Cupola window is of about $72^\circ$. In case of an orbital pass in the zenith, the OGS will be accessible for a time ranging between $50 \sec$ (for $300 km$ ISS altitude) and $70 \sec$ (for $400km$ ISS altitude). Generally, depending on the actual elevation angle of the ISS the actual access time is further limited. Assuming the OGS Lustb\"uhel in Graz (Austria) the ISS would be accessible for more than $20\sec$ for seven passes in October $2012$ at night time. Note, that according to the link model, $\approx10$ $\sec$ are sufficient to collect enough data for the experiment. Yaw and roll movements of the Station will slightly change link-times but this issue will be subjected to a more detailed study in the future.}
                \label{geometry}
\end{figure}

Previously proposed scenarios for quantum experiments using satellites considered down-links to optical ground stations \cite{Kaltenbaek03}. This has the advantage that the channel loss is reduced compared to an uplink scenario, due to the shower-curtain effect \cite{Dror98} caused by the atmosphere. The disadvantage is that the complex quantum sources have to operate in space, although many of the required components (lasers, crystals,...) do not yet have a space-qualification. In an uplink scenario, which is the subject of this paper, the complex and non-space-qualified sources remain on the ground. However, the higher channel loss can be tolerated using state-of-the-art entangled photon sources (EPS) \cite{wittmann12} and faint laser pulse sources (FPS) \cite{jennewein11,marc11}. Already space-qualified single-photon detectors \cite{sun04} will be implemented in space, as they are already a well established technology.

\begin{figure}[ht]
                \centering
  \includegraphics[width=0.49\textwidth]{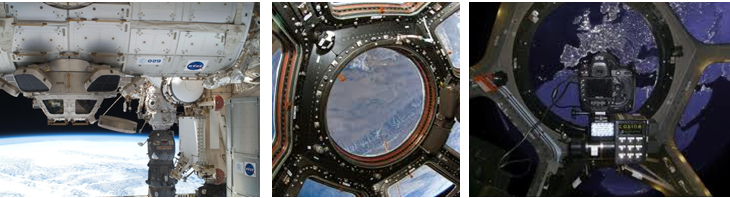}
                \caption{The Cupola Module (left) features a $70.6$ cm window (middle) oriented to the ISS' NADIR. The optical receiver is proposed to be implemented in the focal plane of an existing $D_R=14.3$ cm clear aperture photographer's lens. This lens is mounted on a motorized camera pod, also called NightPOD (right), which is capable of tracking ground targets for up to $70$ seconds. (Picture courtesy of ESA).}
                \label{cupola}
\end{figure}

We propose to use an optical ground station (OGS) as a transmitter for sending one photon of an entangled pair, or alternatively faint laser pulses, to the ISS (see Figure \ref{geometry}). The space station consists of several manned modules, one of which is the so-called Cupola Module (see Figure \ref{cupola} left) and features a circular NADIR facing glass window $70.6$ cm in diameter (see Figure \ref{cupola} middle). A photographer's lens with a clear aperture of $D_R=14.3$ cm has been used to take pictures of ground targets with integration times of up to ten seconds. In order to compensate for the orbital movement, a motorized camera pod (see Figure \ref{cupola} right) was developed and launched in February 2012. A dedicated photon detection module (including the polarization analysis, electronics, storage and communication capabilities) will have to be developed, launched to the ISS, and attached to the photographer's lens replacing the camera presently in use.

As a first step, we propose conducting a Bell-type \cite{Bell64a} experiment between ground and space in the form of a test of the CHSH-type Bell inequality \cite{Clauser69a}. This inequality states that under the assumptions of locality and realism, the strength of correlations between dichotomic polarization measurement outcomes can not lead to a so-called Bell parameter greater than $2$. Considering entangled states, quantum mechanics, however, predicts for a certain set of measurement settings, a maximal Bell value of $2\cdot \sqrt{2}\approx2.8$. This leads to a theoretical contradiction between the predictions of quantum mechanics and classical physics, which can be experimentally tested. In a second series of experiments, the generation of a secret key based on the distribution of entangled photon pairs \cite{Ekert91}, as well as a decoy-state BB84 protocol \cite{Bennett84,hwang03,wang05,Lo05} with faint laser pulses is foreseen. As will be discussed later, these experiments are possible within a few orbital passes of the International Space Station over the transmitter station with the setup sketched in Figure \ref{geometry}. Continuative experiments like quantum teleportation \cite{Ma12b, Ma12a}, entanglement assisted clock synchronization \cite{Valencia02} and also LIDAR \cite{wandlinger05} experiments at very low light levels are conceivable using the same configuration.

In  order to assess the feasibility of the proposed experiments, we first need to define the requirements for their successful demonstration. Measurement errors arise from polarization mismatch between the ground- and the flight-segment and noise events caused by intrinsic detector dark counts, eventual light sources within the detectors field-of-view (FOV) as well as by multi-pair emissions of the quantum source \footnote{Multi-pair emission is only an issue for EPS sources.}, limiting the obtainable signal-to-noise ratio (SNR). The minimum SNR required to prove the presence of entanglement in a Bell-type experiment is $\textrm{SNR}_{min}=2/(\sqrt{2}-1)\approx4.8$. In a QKD experiment, measurement errors lead to bit errors in the generated key, which have to be corrected using classical post-processing algorithms. Yet, a secret key can only be distilled if the quantum bit error ratio (QBER) is below $11$\% \cite{Shor00a}, corresponding to a $\textrm{SNR}>9$. Consequently, the secret key rate also depends on the experimentally obtainable SNR. Note that, in a QKD experiment the minimal SNR needs to be higher than in a Bell experiment, as well as the minimum number of measured events. The latter is on the order of $10^4$ in a QKD experiment \cite{renner11a} and only about $10^3$ for violating a CHSH-inequality with a statistical significance of more than 3 standard deviations.
\begin{figure}[ht]
                \centering
\includegraphics[width=0.49\textwidth]{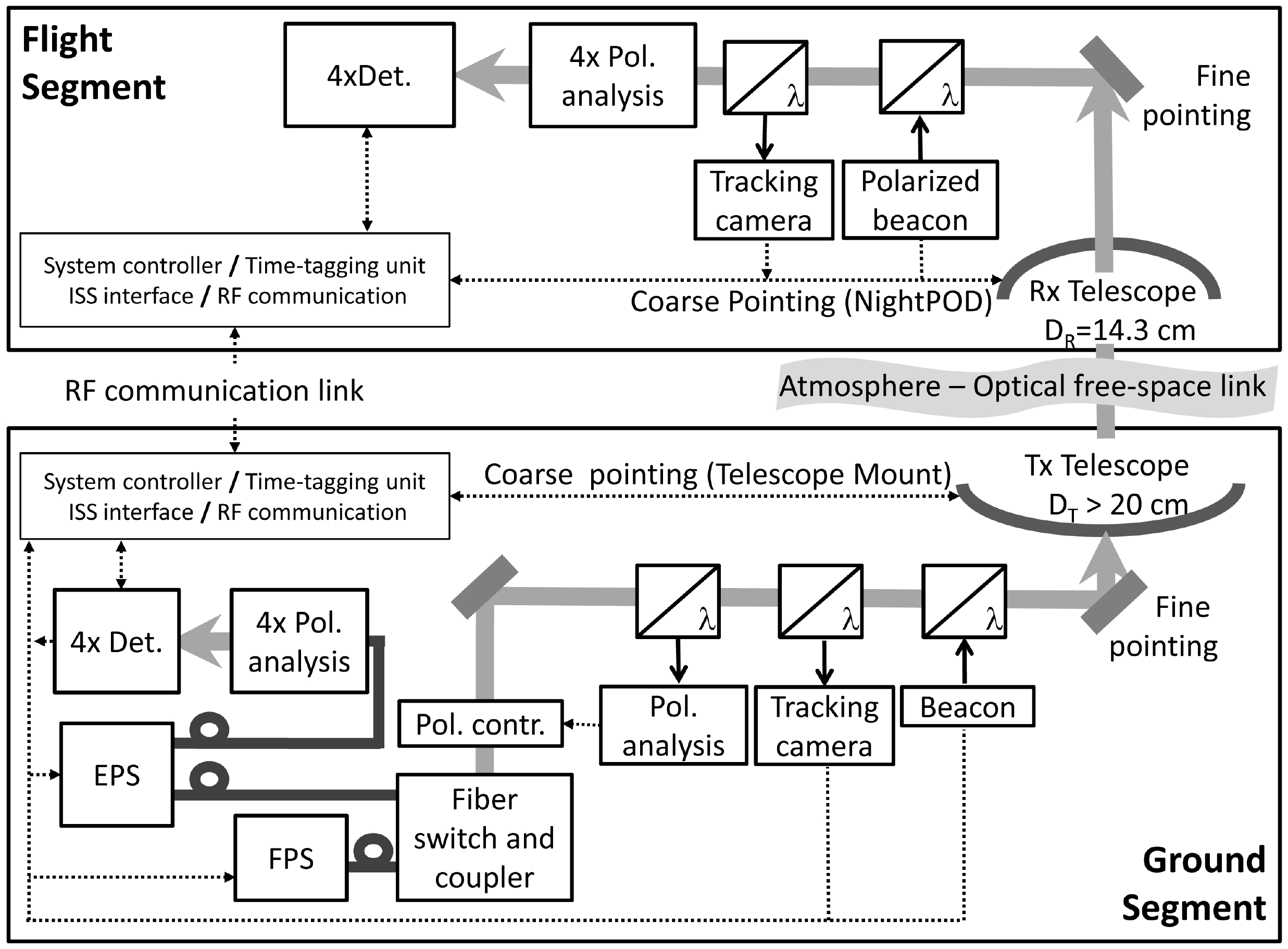}
                \caption{Block diagram of the ground and flight segment, including the entangled photon source (EPS), faint laser pulse source (FPS), polarization controller (Pol. contr.), half-wave plate (H), beam-splitter (B), polarizing beam splitters (P), radio-frequency (RF). For a detailed explanation please refer to the main text.} %Photons from the faint laser pulse source (FPS) or alternatively from the entangled photon source (EPS) are coupled to the optical path of the OGS and are guided via a point-ahead and tip/tilt mirror (fine pointing) to the motorized telescope's (coarse pointing) aperture for being sent towards the International Space Space (ISS). Polarization is compensated for one respective ISS orbit before the quantum link is established, using the polarization controller (pol. contrl.). Beacon light sources, operating at a different wavelength as the quantum singal, might be used fine pointing and tracking purposes during data acquisition. Four single photon detectors are placed behind a four channel polarization analyzer module consisting of half-wave plates (H), beam-splitters (B) and polarizing beam splitters (P). Radio-frequency (RF) based communication links are used for classical communication between the ground- and the flight-segment.}
                \label{block}
\end{figure}

\textit{The transmitter:} An EPS and a FPS will be implemented at an optical ground station. The EPS will emit two photons, each coupled into a separate optical single mode fiber. One photon will be analyzed and detected locally, its twin will be sent to the ISS. As a consequence of unavoidable losses already within the entangled photon source, only a fraction of the generated photon pairs are detectable. This coupling efficiency is typically in the order of $50$\% and already includes losses on optical surfaces, fiber coupling and detection \cite{wittmann12}. Specifically, from $20$ Mcps pairs generated in the crystal, $10$ Mcps single photons in each arm and approx. $5$ Mcps entangled pairs can be detected directly at the source. A typical decoy-state FPS on the other hand operates at a repetition rate of $100$ MHz \cite{Schmitt07,marc11} emitting a mean single photon number per pulse of up to $\mu=1$. Here, the photons are already available in a single mode fiber such that no additional coupling losses have to be considered. This results in approximately $40$ Mcps detection events locally, assuming a detection efficiency of $50$\%.

As depicted in Figure \ref{block}, photons of either source are coupled out of the fiber to the optical path of the OGS and are then guided via a point-ahead and tip/tilt mirror (fine pointing) to the motorized telescope's (coarse pointing) sending aperture.  Note that pointing, acquisition and tracking of low-earth-orbiting (LEO) satellites is a well-established technology \cite{Toyoshima07b,Giggenbach07}. Additionally, a polarization compensator is used at the ground station to establish a common polarization reference frame between the ground- and the flight-segment. Time-tagging units will be used to store the timing information of the local events for later analysis at the optical ground station. In the EPS case, the local events are the detection times of the locally measured photons, while for the FPS, these are the laser trigger signals.
\begin{figure}[ht]
       %         \centering
  \includegraphics[width=0.48\textwidth]{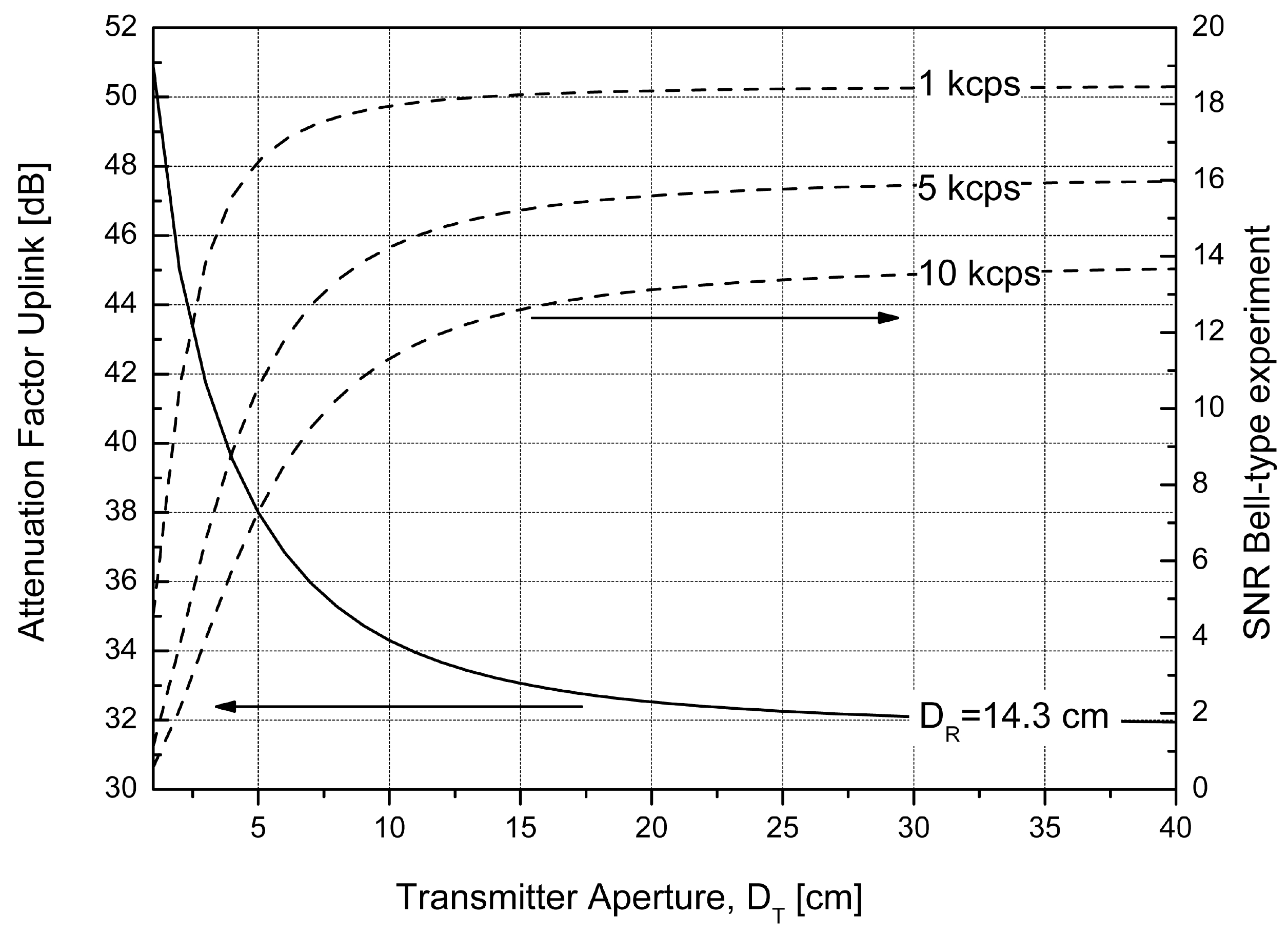}
                \caption{The graph shows the attenuation of the uplink versus the transmitter aperture $D_{T}$ for an onboard available receiving lens with a clear aperture of $D_{R}=14.3$ cm (solid line, left axis). It follows that an attenuation factor of $40$ dB has to be expected (including a $5$ dB margin) for the ISS being $400$ km high at zenith. Interestingly, increasing the sending aperture beyond $20$ cm has only a minor effect on the overall link efficiency. This is because diffraction at the sending aperture is dominant for small aperture sizes and fades out as $1/D_{T}^2$. Furthermore, the expected SNR in an Bell-type experiment is shown for three different background levels (dashed line, right axis).}
                \label{snr}
\end{figure}

\textit{The uplink:} A calculation of the attenuation factor for an optical free-space link depending on the diameters of the transmitter $D_T$ and the receiver $D_R$ apertures is given in Ref. \cite{klein74}, which also includes pointing errors, optical losses and realistic atmospheric turbulence. Figure \ref{snr} shows a plot of the attenuation factor as a function of the on-ground $D_T$ for the lens available onboard the ISS with a clear aperture of $14.3$ cm. Assuming the photographer's lens has a clear aperture of $14.3$ cm, we expect the total link attenuation to be $40$ dB. This includes a $20$ cm transmitting aperture and an additional $\approx 5$ dB margin for effects, not included in the calculation. This result agrees very well with an independent calculation given in \cite{jennewein11}. Note that, most of the optical ground stations feature even larger sending apertures than $20$ cm.

\textit{The receiver:} The motorized camera pod, also called NightPOD, is able to tilt the photographer's lens over a total angle of $ 72^\circ$ from one window edge to the other \footnote{For safety reasons, the NightPOD actually tilts for a total angle of $36^\circ$ only.}. This puts a restriction on the minimal elevation angle the ISS needs to be before quantum communication can commence. Depending on the actual roll/yaw/tip angles, this is on the order of $51^\circ$ and enables quantum communication for a duration of up to $70$ seconds per orbital pass. Furthermore, some 7 to 8 passes of the ISS at nighttime over an OGS in Tenerife (Spain) or Graz (Austria) are to be expected per month. Since the window consists of four vacuum/air spaced layers of broad band coated BK7 glass with a total thickness of $20.8$ cm, we are aware of possible polarization effects for flat incident angles of the photons. However, we expect these effects to be negligible for deviations less than $\pm 10^\circ$ from normal incident, such that the usable link-time per orbital pass would be reduced to about $20$ seconds in the worst case. Because of this, the number of useful orbital passes would also be reduced.

For maintaining the optical link with an OGS, coarse pointing with a pointing error less than $10^{\prime\prime}\approx 50$ $\mu$rad is already archived by the NightPOD motor. Commercially available single photon detection modules for measuring the impinging photons use silicon based avalanche photo-diodes (Si-APD) with an active area of up to $500$ $\mu$m. These modules will be placed after the polarization analysis in the focal plane of the $f=400$ mm lens, which corresponds to a field-of-view (FOV) of approximately $1$ mrad. Hence, the detectors visible footprint diameter at ground is approximately $400$ m for the ISS (at $400$ km height) being at zenith and approximately $500$ m for the ISS being at an elevation angle of $51^\circ$. Based on our experiences from ground based experiments so far \cite{ursin07,Schmitt07}, we expect background counts from light sources within that FOV to be on the order of $100$ cps, depending on the actual location of the OGS. For example, the OGS of the European Space Agency in Tenerife (Spain) is located far outside any artificial light sources at an astronomical site. The coarse pointing capabilities of the NightPOD is sufficient to keep the impinging single photons on the detector's active area. Nevertheless, a fine pointing facility (tip/tilt mirror or the NightPOD motorized axis) together with a CCD camera might be necessary for pointing and tracking purposes to compensate vibrational effects of the space station (see Figure \ref{block}). The initial pointing, will require the ISS orbit telemetry data available to the NightPOD.

The envisaged dedicated quantum payload consists of a $50:50$ beam-splitter, a half-wave plate, and polarizing beam-splitters, such that the polarization of the photons is randomly analyzed either in the $H/V$ or the $\pm45^\circ$ basis. An extra half-wave-plate at the entrance of the detection module enables measurement in another pair of complementary linear polarization bases, as is required by the CHSH-inequality. During data accumulation, a time-tagging unit stores the timing and detector-channel information to a local storage drive for transmission to the ground via RF at some later time. In order to find the associated detection events between ground and space, the cross-correlation of the individual time-tag data sets is calculated. In a QKD experiment, the remaining errors in the quantum key are eliminated during post processing using classical algorithms. These algorithms require classical communication between the satellite and the ground and can be executed directly with the time-tagging units.

\begin{figure}[ht]
            %    \centering
  \includegraphics[width=0.48\textwidth]{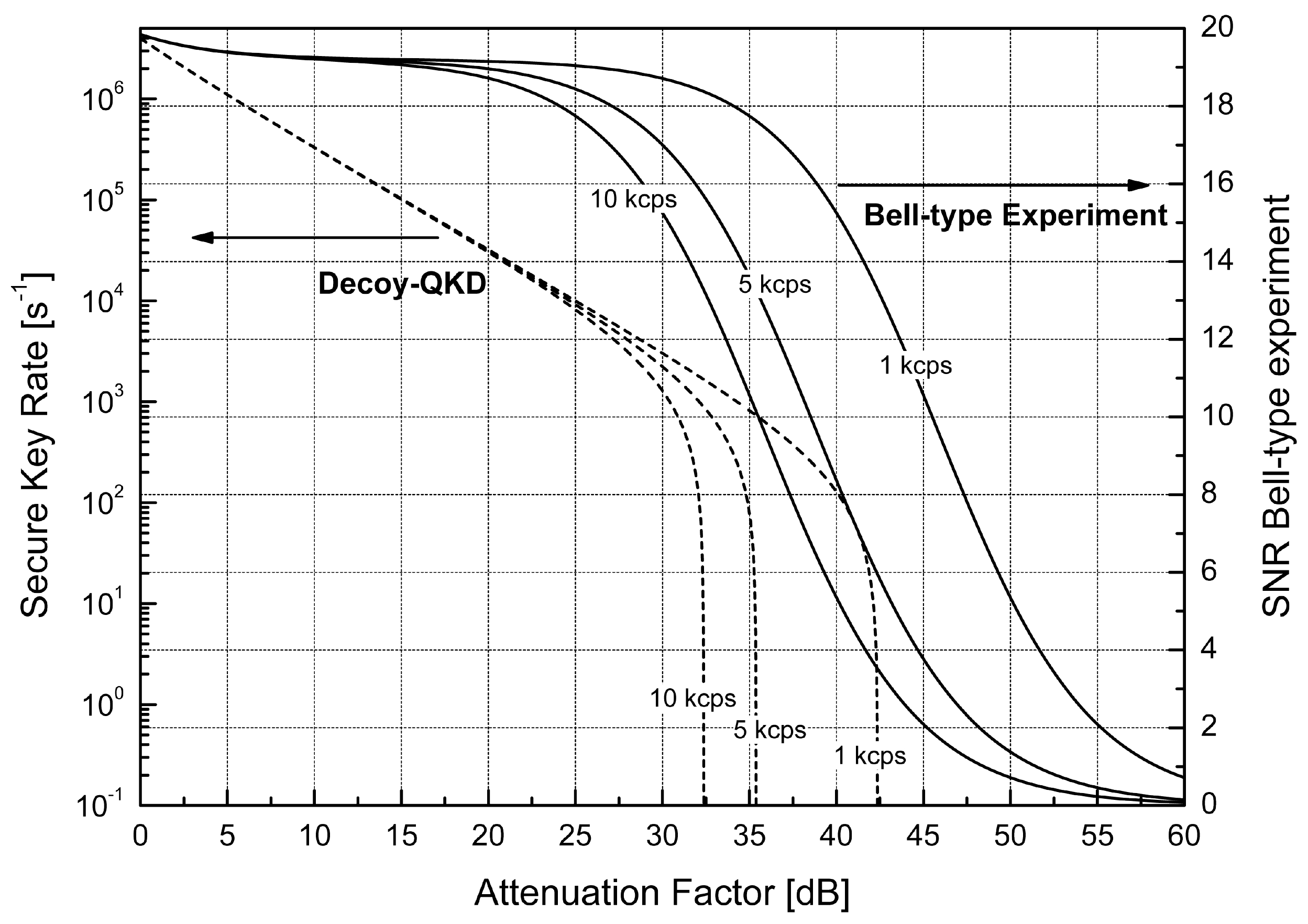}
                \caption{This graph shows the obtainable SNR in a Bell experiment as a function of the attenuation factor for different background levels (slid lines, right axis). Additionally, the expected secure key rate in a FPS decoy-state protocol is depicted (dashed lines, left axis). The experimental parameters of the EPS, the FPS and the single photon detectors are described in the main text.}
                \label{visi}
\end{figure}

\textit{Experiments:} As discussed above, the loss over the optical uplink from the OGS to the ISS will be round $40$ dB. Hence, for a Bell-type experiment we expect $1$ kcps single photons from the source being detected, corresponding to $500$ cps detected entangled photon pairs. For quantum key distribution based on a FPS decoy-state source, we expect $4$ kcps detection events at the space station. We study the experimental imperfections mentioned earlier in more detail and base our considerations on a theoretical model devised in \cite{xma07}. Figure \ref{visi} shows the SNR of a Bell-type experiment, and the secret key rate in a QKD experiment, as a function of the attenuation through the uplink from the OGS to the ISS, for different levels of background counts in the space-based detectors. From the graph we see that, for a Bell-type experiment with a link attenuation as expected for the herein proposed mission concept, a SNR of up to $1:15$ can be achieved with a realistic background count rate of $1$ kcps. As mentioned earlier, the quantum efficiency of the single photon detectors are approximately $50$\% with a dark count rate of about $500$ cps. We are aware of radiation effects typically increasing the dark counts \cite{sun01}, but this can be mitigated by proper shielding. In fact, the background counts can go up to about $10$ kcps, leading to a SNR of $1:5$, which is still above the limit for a violation of the CHSH inequality. This margin, with respect to the tolerable background rate, would support the scientific claim to have distributed entanglement between the ground- and flight-segment. Regarding the distribution of a secure quantum key, the situation becomes critical at higher background levels. Yet, assuming $1$ kcps background counts, the successful demonstration of quantum key distribution seems to be possible, too.

As discussed above, the quantum link can be maintained for up to $20-70$ $\sec$ within one orbital pass (see Fig. \ref{geometry}). Hence, the scientific goal of violating a Bell-inequality by $3$ standard deviations of statistical significance is possible within one satellite pass, since more than $10^3$ coincidences will then have been identified. The same is true for the QKD experiment based on the FPS decoy-state protocol. In this case, more than $10^4$ events \cite{renner11a} would have been collected after one orbital pass.

We have shown that by using existing infrastructure inside the ISS, one can perform quantum communication experiments, by adding only a scientific payload the size of a photographer's camera (see Figure \ref{block}). Both major scientific goals could be achieved with only a few orbital passes of the ISS over the OGS. A successful demonstration of these experiments will provide the basis for a whole variety of additional future experiments (e.g. quantum communication in a down-link or even an inter-satellite link scenario) and will prove the feasibility of global quantum communication using state-of-the-art technology as a kind of path-finder mission.

Special thanks to cosine B.V. for the development of the incredible NightPOD under GSTP funding from the European Space Agency (ESA) and for the open discussion on its specifications mentioned in this text. The authors thank Anton Zeilinger and Peev Momshil for discussions.

\bibliographystyle{abbrev}

%\bibliography{../Ursin}
%%\bibliographystyle{plain}
%\bibliographystyle{unsrt}

\end{document}